\documentclass[conference,twoside,letterpaper,10pt]{IEEEtran}
\IEEEoverridecommandlockouts

\ifCLASSINFOpdf
\else
\fi

\usepackage[cmex10]{amsmath}
\usepackage{cite}
\usepackage{mathtools}
\usepackage{flushend}
\usepackage{url}
\usepackage{enumerate}
\usepackage{comment}
\begin{document}
%
\title{Impact of Received Signal on Self-interference Channel Estimation and Achievable Rates in In-band Full-duplex Transceivers \vspace{-0mm}}

\author{\IEEEauthorblockN{Dani~Korpi,
Lauri~Anttila,
and~Mikko~Valkama}
\\
\IEEEauthorblockA{Department of Electronics and Communications Engineering, Tampere University of Technology, Finland\\ e-mail: dani.korpi@tut.fi, lauri.anttila@tut.fi, mikko.e.valkama@tut.fi\vspace{-0mm}}
\thanks{The research work leading to these results was funded by the Academy of Finland (under the project \#259915 "In-band Full-Duplex MIMO Transmission: A Breakthrough to High-Speed Low-Latency Mobile Networks"), the Finnish Funding Agency for Technology and Innovation (Tekes, under the project "Full-Duplex Cognitive Radio"), the Linz Center of Mechatronics (LCM) in the framework of the Austrian COMET-K2 programme, and Emil Aaltonen Foundation.}}%

\maketitle

\begin{abstract}

In this paper we analyze the effect of the calibration period, or lack of, on self-interference channel estimation in the digital domain of in-band full-duplex radio transceivers. In particular, we consider a scenario where the channel estimation must be performed without a separate calibration period, which means that the received signal of interest will act as an additional noise source from the estimation perspective. We will explicitly analyze its effect, and quantify the increase in the parameter estimation variance, or sample size, if similar accuracy for the channel estimate is to be achieved as with a separate calibration period. In addition, we will analyze how the calibration period, or its absence, affects the overall achievable rates. Full waveform simulations are then used to determine the validity of the obtained results, as well as provide numerical results regarding the achievable rates. It is shown that, even though a substantial increase in the parameter sample size is required if there is no calibration period, the achievable rates are still comparable for the two scenarios.


\end{abstract}

\section{Introduction}

Full-duplex radio communications where transmission and reception are done simultaneously at the same radio frequency (RF) carrier has recently gained considerable interest among researchers. In theory, it has the potential to even double the spectral efficiency of current communication systems. This makes it an appealing concept when trying to increase the data rates of the individual users to match the ever increasing demands. There have already been several promising demonstration-type implementations of such full-duplex radio transceivers \cite{Knox12,Choi10,Jain11,Duarte10,Bharadia13}. There is also a wide body of literature regarding the theoretical analysis of in-band full-duplex communications under various circuit impairments and deployment scenarios \cite{Day12,Korpi133,Korpi13,Syrjala13,Riihonen122,Ahmed133}.

The fundamental challenge behind in-band full-duplex communications is the problem of the own transmit signal coupling back to the receiver. This so-called \textit{self-interference} (SI) must be heavily attenuated, as it will otherwise saturate the receiver chain, or in the very least make the detection of the received signal of interest very challenging. Typically, the SI signal is first attenuated at the input of the receiver chain by subtracting a properly delayed and attenuated version of the own transmit signal from the total received signal \cite{Choi10,Jain11,Duarte10}. This cancellation stage is referred to as RF cancellation, and it decreases the power of the total receiver input signal to a suitable level so that the receiver chain components will not be completely saturated. Usually, additional SI cancellation is still performed in the digital domain, referred to as digital cancellation \cite{Jain11,Anttila13}. Ideally, at this point the SI signal is attenuated sufficiently low to achieve an adequate signal-to-interference-plus-noise ratio (SINR) for detecting the received signal of interest.

A crucial step in in-band full-duplex communications is the estimation of the SI channel. A common assumption in earlier literature has been that there is a \textit{separate calibration period} for this purpose, during which only the device itself is transmitting \cite{Duarte12,Korpi133,Jain11,Aggarwal12,Duarte14}. Thus, only the SI signal is being received, as there is no signal of interest present, allowing for accurate estimation of the SI channel. However, the downside of this approach is the need for a reception-free period each time the SI channel must be estimated. This obviously decreases the achievable total data rate, as the two communicating parties must revert back to half-duplex mode during each calibration period. Furthermore, especially in mobile devices, the SI channel characteristics are time-varying \cite{Jain11}, and thus periodically repeating calibration is needed.

In this paper, we compare the above situation to a scenario where there is no such calibration period, which means that the SI channel must be estimated when also the received signal of interest is included in the total signal. Thus, the received signal of interest acts as an additional noise source from the estimation perspective. In particular, we concentrate on the channel estimation in the digital domain, where the SI signal has already been attenuated by RF cancellation. A more detailed analysis regarding the channel estimation for RF cancellation is left for future work. In the digital domain, increased noise level during the SI channel estimation increases the parameter estimation variance, or the corresponding sample size required to achieve a certain level of SI cancellation. Below, we will quantify this, and provide an equation for the estimation sample sizes with and without a calibration period, when the same accuracy for the SI channel estimate is required in both cases. Furthermore, all the derivations are done using a general MIMO full-duplex signal model. Although there have already been studies where the SI channel estimation is performed in the presence of the signal of interest \cite{Li11,Choi10}, to the best of our knowledge this is the first time the accuracy of the estimate is analytically quantified in such a scenario.

In addition, we perform an analysis into the achievable rates with and without a separate calibration period. Namely, it will be determined if it is preferable to have the additional overhead introduced by the calibration period, or a less accurate SI channel estimate. In the former case, the achievable rate is decreased by the reception-free period during SI channel estimation, and in the latter the more powerful residual SI is limiting the final SINR. Determining whether a separate calibration period is beneficial or not is a crucial design problem for any network utilizing in-band full-duplex transceivers and it depends heavily on the coherence time of the SI channel. This is due to the fact that, for a shorter coherence time, the SI channel estimation must be repeated more frequently. To the best of our knowledge, this issue has not been addressed in earlier literature.

The rest of this paper is organized as follows. In Section~\ref{sec:est_var} we derive a relation between the parameter estimation sample sizes for scenarios with and without a calibration period, to achieve a given estimator variance. After this, in Section~\ref{sec:ach_rates}, expressions for the achievable rates with and without a calibration period are derived. Then, in Section~\ref{sec:simul}, the obtained theoretical results are compared with the results from full waveform simulations, and numerical values for the achievable rates are provided. Finally, the conclusions are drawn in Section~\ref{sec:conc}.




\section{Lower Bound for Channel Estimator Variance}
\label{sec:est_var}

In this paper we consider a block wise estimation procedure, where a certain number of samples is used to obtain a SI channel estimate for a time period of specific length. It is assumed that this estimate is valid for the duration of the SI channel coherence time, after which it must be estimated again \cite{Jain11}. 

The quality of the SI channel estimate is assessed by determining the lower bound of the achievable variance for the estimate with and without a separate calibration period. This will allow the comparison of the two cases in terms of the required parameter estimation sample size, which reveals the increase in computational complexity caused by the presence of the received signal of interest. In practice, the derived lower bound for the variance might not be achieved, but as we are only interested in the relation of the two cases, the absolute values of the obtained variances are not crucial. Furthermore, we will show later with full waveform simulations, that the chosen approach, utilizing lower bounds for the variances, does in fact produce reliable results under practical circumstances.

The analysis in this paper covers two scenarios: a linear signal model and a widely-linear signal model. In the former, it is assumed that the SI signal experiences only linear distortion, and thereby it can be efficiently cancelled with linear processing. However, it has been observed in earlier literature that typically linear modeling of the SI channel is not accurate, as different analog impairments distort the SI signal in numerous ways \cite{Korpi13,Korpi133}. For this reason, we will also perform the analysis with widely-linear signal model and widely-linear digital SI cancellation \cite{Korpi133}. It has been shown that the widely-linear signal model is valid up to transmit powers of roughly 15 dBm, depending obviously on the exact component parameters. Thus, in addition to the linear model, we will extend the widely-linear signal model presented in \cite{Korpi133} to a MIMO scenario, and analyze it as well.

\subsection{Linear MIMO Signal Model}
\label{sec:linear_model}

Let us first consider a linear signal model for a MIMO full-duplex transceiver. Then, the received signal after the analog-to-digital converter (ADC) of the $i$th receiver chain can be written as
\begin{align}
	y_{i,ADC}(n) = \sum_{j=1}^{N_{tx}} h_{ij}(n) \star x_{j,ref}(n) + r_{i}(n) + z_i(n) \text{,} \label{eq:y_adc_n}
\end{align}
where $N_{tx}$ is the number of transmitters, $r_i(n)$ is the received signal of interest after amplification and digitizing, $z_i(n)$ represents additional noise sources, $x_{j,ref}(n)$ is the $j$th transmitted signal at the point where the reference signal for digital cancellation is taken, and $h_{ij}(n)$ is the effective channel experienced by $x_{j,ref}(n)$ when propagating to the $i$th receiver baseband. The signal model has been written in terms of the reference signals as they are obviously known in the receiver, and the SI channel estimation is thereby performed with respect to them. The most typical selection for the digital cancellation reference signals is to use the original transmitted samples, in which case $x_{j,ref}(n) = x_j(n)$, where $x_j(n)$ is the $j$th original transmitted waveform. However, for generality, the utilized signal model does not assume anything regarding the reference signal, and thus the obtained results can be applied for various scenarios.

To allow for a more expressive analysis, let us write the signal model in \eqref{eq:y_adc_n} with vector notation over some observation period as follows:
\begin{align}
	\mathbf{y}_{i,ADC} = \sum_{j=1}^{N_{tx}} \mathbf{X}_{j,ref} \mathbf{h}_{ij} + \mathbf{r}_{i} + \mathbf{z}_i\text{,} \label{eq:y_adc}
\end{align}
where $\mathbf{X}_{j,ref}$ is a covariance windowed convolution matrix of the form
\begin{align*}
\mathbf{X}_{j,ref} =  \left[\begin{smallmatrix}
  x_{j,ref}(M-1) & x_{j,ref}(M-2) & \cdots & x_{j,ref}(0) \\
  x_{j,ref}(M) & x_{j,ref}(M-1) & \cdots & x_{j,ref}(1) \\
  \vdots  & \vdots  & \ddots & \vdots  \\
  x_{j,ref}(N-1)& x_{j,ref}(N-2) & \cdots & x_{j,ref}(N-M)
 \end{smallmatrix}\right]
\end{align*}
and $\mathbf{y}_{i,ADC}$ is of the form
\begin{align*}
	\mathbf{y}_{i,ADC} = \left[\begin{smallmatrix} y_{i,ADC}(M-1) & y_{i,ADC}(M) & \cdots &  y_{i,ADC}(N-1) \end{smallmatrix}\right]^{T} \text{.}
\end{align*}
Here, $N$ is the parameter estimation sample size and $M$ is the length of an individual channel response estimate. The vector-based signal model in \eqref{eq:y_adc} can be further simplified by denoting
\begin{align*}
&\mathbf{X}_{ref} = \begin{bmatrix} \mathbf{X}_{1,ref} & \mathbf{X}_{2,ref} & \cdots &  \mathbf{X}_{N_{tx},ref} \end{bmatrix} \text{, and}\nonumber\\
&\mathbf{h}_i = \begin{bmatrix} \mathbf{h}_{i1}^T & \mathbf{h}_{i2}^T & \cdots &  \mathbf{h}_{iN_{tx}}^T \end{bmatrix}^T \text{.}
\end{align*}
Then, \eqref{eq:y_adc} becomes
\begin{align}
	\mathbf{y}_{i,ADC} = \mathbf{X}_{ref} \mathbf{h}_{i} + \mathbf{r}_{i} + \mathbf{z}_i\text{.} \label{eq:y_adc_s}
\end{align}

In order to be able to attenuate the SI signal, the channel $\mathbf{h}_{i}$ must be estimated. To see how a separate calibration period affects the quality of this channel estimate denoted by $\mathbf{\hat{h}}_i$, let us determine the lower bound for its variance under two circumstances. Namely, when the received signal of interest is included in the total signal during the calculation of $\mathbf{\hat{h}}_i$, and when it is not included in it. First, it is assumed that the noise signal $\mathbf{z}_i$ is Gaussian distributed as $\mathbf{z}_i \sim N(0,\sigma_n^2 \mathbf{I})$, where $\sigma_n^2$ is the power of the thermal noise in the digital domain. In other words, it is assumed that the power of nonlinearities and other sources of distortion is negligibly low.

Under a separate calibration period, and for a given $\mathbf{X}_{ref}$ known inside the device, this ensures that also the signal $\mathbf{y}_{i,ADC}$ is Gaussian distributed. However, it can be shown that this is true also when the received signal of interest is present. Namely, in \cite{Shuangqing10} it is shown that OFDM signals are approximately normally distributed with a sufficiently large number of subcarriers. Assuming further that the consecutive samples of the received waveform are uncorrelated, we have $\mathbf{r}_i \sim N(0,\sigma_r^2 \mathbf{I})$, where $\sigma_r^2$ is the power of the received signal of interest in the digital domain.

Thereby, by treating $\mathbf{X}_{ref}$ as a deterministic matrix, the overall signal $\mathbf{y}_{i,ADC} = \mathbf{X}_{ref} \mathbf{h}_{i}+\mathbf{r}_i +\mathbf{z}_i$ is Gaussian distributed also without a separate calibration period, and in this case its probability density function is as follows \cite{Schreier10}:
\begin{align}
	&p(\mathbf{y}_{i,ADC} | \mathbf{h}_i) = \frac{1}{\sqrt{\pi^{2N}\operatorname{det}(\mathbf{R}_{\mathbf{y}_i| \mathbf{h}_i})}}\nonumber\\
	&\times e^{\left(-\frac{1}{2}\left(\mathbf{y}_{i,ADC}-\mathbf{\mu}_{\mathbf{y}_i|\mathbf{h}_i}\right)^H \mathbf{R}^{-1}_{\mathbf{y}_i| \mathbf{h}_i} \left(\mathbf{y}_{i,ADC}-\mathbf{\mu}_{\mathbf{y}_i|\mathbf{h}_i}\right)\right)} \text{,} \label{eq:y_pdf}
\end{align}
where $\mathbf{R}_{\mathbf{y}_i| \mathbf{h}_i}$ is the augmented conditional covariance matrix of the total received signal, and $\mathbf{\mu}_{\mathbf{y}_i|\mathbf{h}_i}$  is its augmented conditional mean vector. From this, the Cram\'{e}r--Rao lower bound (CRLB) for the channel estimate can be calculated, which can be used to obtain insight into the achievable digital SI cancellation under the two considered scenarios. Based on \eqref{eq:y_pdf}, the CRLB for the channel estimate can be shown to be
\begin{align}
	\operatorname{Cov}(\mathbf{\hat{h}}_i) \geq \mathbf{{X}}^{-1}_{ref} \mathbf{R}_{\mathbf{y}_i| \mathbf{h}_i} (\mathbf{{X}}_{ref}^H)^{-1} \text{.} \label{eq:crlb_01}
\end{align}
Furthermore, it can easily be shown that $\mathbf{R}_{\mathbf{y}_i| \mathbf{h}_i} =(\sigma_n^2+\sigma_r^2)\mathbf{I}$, where $\mathbf{I}$ is an identity matrix. Thus, the CRLB can be rewritten as
\begin{align}
	\operatorname{Cov}(\mathbf{\hat{h}}_i) \geq (\mathbf{{X}}_{ref}^H \mathbf{{X}}_{ref} )^{-1} (\sigma_n^2+\sigma_r^2) \text{.} \label{eq:crlb_1}
\end{align}
The above equation holds for any given $\mathbf{{X}}_{ref}$, and it can be evaluated numerically as long as the elements of the data matrix are known. However, assuming that the consecutive samples of the transmit signals are uncorrelated and that the value of $N$ is large, we can obtain further insight into the estimator variance by approximating the term $\mathbf{{X}}_{ref}^H \mathbf{{X}}_{ref}$ by $N p_{ref} \mathbf{I}$, where $\mathbf{I}$ is an identity matrix and $p_{ref}$ is a constant \cite{Schreier10}. This allows further simplification of \eqref{eq:crlb_1}, and the CRLB for the channel estimate becomes
\begin{align}
	\operatorname{Cov}(\mathbf{\hat{h}}_i) &\geq \left(\frac{\sigma_n^2+\sigma_r^2}{N p_{ref}}\right) \mathbf{I} \label{eq:crlb_2}\text{,}
\end{align}
or for each individual tap of the channel estimate:
\begin{align}
	\operatorname{Var}(\hat{h}_{ij}) &\geq \frac{\sigma_n^2+\sigma_r^2}{N p_{ref}} \label{eq:crlb_3}\text{.}
\end{align}

Assuming a separate calibration period, $\sigma_r^2 = 0$, and the CRLB for each tap becomes
\begin{align}
	\operatorname{Var}(\hat{h}_{ij}) &\geq \frac{\sigma_n^2}{N_c p_{ref}} \label{eq:crlb_c}\text{,}
\end{align}
where $N_c$ is the parameter estimation sample size with a separate calibration period. By requiring similar minimum variance with and without a calibration period, we get then the following relation for the parameter estimation sample sizes:
\begin{align}
	N &= N_c \left(\mathit{snr}+1 \right) \text{,} \label{eq:N_req_lin}
\end{align}
where $\mathit{snr}$ is the signal-of-interest-to-noise ratio for an individual receiver in the digital domain in linear scale. We have briefly presented this result also in an earlier publication \cite{Korpi14}, but here we provide a significant amount of additional details, and we also evaluate it more elaborately. Equation \eqref{eq:N_req_lin} reveals that if, e.g., a SNR of 10~dB is assumed for the received signal of interest, estimating the SI channel without a separate calibration period requires approximately 11 times more samples if a similar accuracy is to be achieved as with a calibration period. Also note that, even though the derivation of \eqref{eq:N_req_lin} is done by comparing the CRLBs for the two different scenarios, we will show later in this article with full waveform simulations that it does indeed provide reliable information with practical estimators, such as least squares, and even when the considered MIMO full-duplex transceiver is not ideal in terms of its RF components.


\subsection{Widely-Linear MIMO Signal Model}
\label{sec:wl_model}

In \cite{Korpi133} it is shown that the IQ imaging occurring in a full-duplex transceiver is under realistic circumstances the most dominant source of distortion after linear SI. There, the authors proposed a novel widely-linear digital cancellation algorithm that is capable of modeling also the IQ imaging, and it was demonstrated that it provides a substantial improvement in the achievable SINR. For this reason, we will also determine the effect of a separate calibration period using a widely-linear signal model, which considers the effect of practical IQ imaging.

\begin{figure*}[!t]
\centering
\includegraphics[width=\textwidth]{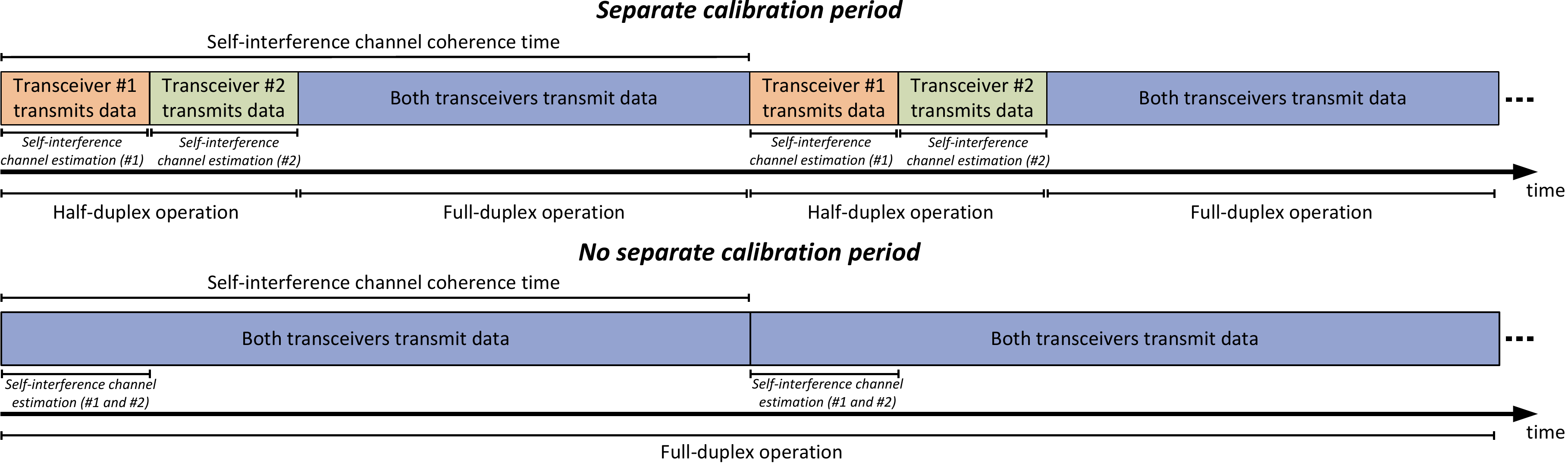}
\caption{A timeline of the proposed procedure for SI channel estimation with and without periodic calibration periods for two communicating in-band full-duplex transceivers.}
\label{fig:mac}
\vspace{-3mm}
\end{figure*}

Using the same notation as in \eqref{eq:y_adc_n}, let us now rewrite the digitized $i$th received signal as follows:
\begin{align}
	y_{i,ADC}(n) &= \sum_{j=1}^{N_{tx}} \left(h_{ij,1}(n) \star x_{j,ref}(n)\right. \nonumber\\
&\left. {} + h_{ij,2}(n) \star x^{\ast}_{j,ref}(n)  \right)+ r_{i}(n) + z_i(n) \text{,} \label{eq:y_adc_wl_n}
\end{align}
where $h_{ij,1}(n)$ and $h_{ij,2}(n)$ are the responses for the direct and image components, respectively, and $()^{\ast}$ denotes complex conjugation.

Similar to the linear signal model, \eqref{eq:y_adc_wl_n} can be also expressed with concise vector notation as:
\begin{align}
	\mathbf{y}_{i,ADC} = \mathbf{X}_{ref,WL} \mathbf{h}_{i,WL} + \mathbf{r}_{i} + \mathbf{z}_i\text{,} \label{eq:y_adc_wl_si}
\end{align}
where 
\begin{align*}
&\mathbf{X}_{ref,WL} = \left[\begin{smallmatrix} \mathbf{X}_{1,ref} & \mathbf{X}^{\ast}_{1,ref} & \mathbf{X}_{2,ref} & \mathbf{X}^{\ast}_{2,ref} & \cdots &  \mathbf{X}_{N_{tx},ref}^{\ast} \end{smallmatrix}\right] \text{, and}\nonumber\\
&\mathbf{h}_{i,WL} = \begin{bmatrix} \mathbf{h}_{i1,1}^T & \mathbf{h}_{i1,2}^T & \mathbf{h}_{i2,1}^T & \mathbf{h}_{i2,2}^T & \cdots &  \mathbf{h}_{iN_{tx},2}^T \end{bmatrix}^T \text{.}
\end{align*}
Based on \eqref{eq:y_adc_wl_si}, it is then possible to obtain an estimate for $\mathbf{h}_{i,WL}$, similar to the linear signal model. In addition, since the structure of the received signal is still similar to that presented in Section~\ref{sec:linear_model}, the CRLB for the channel estimate can be written similarly as
\begin{align}
	\operatorname{Cov}(\mathbf{\hat{h}}_{i,WL}) \geq (\mathbf{{X}}_{ref,WL}^H \mathbf{{X}}_{ref,WL} )^{-1} (\sigma_n^2+\sigma_r^2) \text{.} \label{eq:crlb_wl_1}
\end{align}

Assuming that the realizations for the reference signals are drawn from circular data, we can again approximate $\mathbf{{X}}_{ref,WL}^H \mathbf{{X}}_{ref,WL}$ by $ N \approx p_{ref,WL} \mathbf{I}$, where $p_{ref,WL}$ is a constant \cite{Schreier10}. Note that usually communication signals fulfill the circularity criterion, and thereby this assumption does not diminish the applicability of the obtained results, as long as the reference signals are now taken from the original transmit samples. Because of this, the CRLB for the widely-linear signal model is identical to the CRLB for the linear model, up to a scaling constant. This means that \eqref{eq:N_req_lin} applies also in this case.


\section{Achievable Rates with and without a Separate Calibration Period}
\label{sec:ach_rates}



As discussed earlier, the inclusion of a calibration period is a fundamental problem in networks based on in-band full-duplex transceivers. To provide some insight into the benefits and drawbacks of having a separate calibration period, we will now analyze the achievable rates with and without it.

First, let us write the signal after digital SI cancellation. Assuming linear cancellation processing, the signal at the input of the detector is as follows:
\begin{align}
	\mathbf{y}_{i} &= \mathbf{y}_{i,ADC}-\mathbf{X}_{ref} \mathbf{\hat{h}}_{i} \nonumber\\
	&= \mathbf{X}_{ref} \left(\mathbf{h}_{i} - \mathbf{\hat{h}}_{i}\right) + \mathbf{r}_{i} + \mathbf{z}_i \text{,} \label{eq:y_adc_dc}
\end{align}
where $\mathbf{\hat{h}}_{i}$ is the linear channel estimate. The SINR after digital cancellation is directly defined as the ratio of the power of the received signal of interest, written as the diagonal elements of $\operatorname{Cov}\left(\mathbf{r}_{i}\right)$, to the power of the noise and residual SI, written as the diagonal elements of $\operatorname{Cov}\left(\mathbf{X}_{ref}\left(\mathbf{h}_{i} - \mathbf{\hat{h}}_{i}\right) + \mathbf{z}_i\right)$.

Without a calibration period, the SI channel is estimated during normal full-duplex operation, i.e., when also the received signal of interest is included in the signal, as illustrated in the lower part of Fig.~\ref{fig:mac}. This means that the achievable instantaneous rate (per channel use) for an individual spatial data stream\footnote{In the final version of the paper we will provide these expressions also for a MIMO system utilizing spatial multiplexing.} can be written as
\begin{align}
	C_{nc} = 2\log_2\left(1+\mathit{sinr}_{nc}\right) \text{,} \label{eq:cap_nc}
\end{align}
where $\mathit{sinr}_{nc}$ is the achieved instantaneous SINR at detector input without a calibration period. The rate is multiplied by $2$ as we consider the overall achievable rate between two communicating parties. It should be noted that the above expression assumes the noise and residual SI to be Gaussian distributed. Strictly speaking, this is of course not the case, but this assumption allows us to determine the achievable instantaneous rate in closed form. It is also a typical assumption in the literature and has been observed to provide realistic results \cite{Day12,Ahmed133,Aggarwal12}.

In the scenario where a separate calibration period for SI channel estimation is assumed, some of the rate is lost due to reverting back to half-duplex mode. The upper part of Fig.~\ref{fig:mac} illustrates a possible procedure for SI channel estimation using a calibration period, when two full-duplex transceivers are communicating. Effectively, the SI channel estimation is performed by transmitting data in half-duplex mode for a certain period of time, during which both of the transceivers estimate their SI channels in turn. This must be done repeatedly between certain intervals, based on the coherence time of the SI channel. In this paper we assume that it is sufficient to estimate the SI channel once during the coherence time. One immediate observation is that more of the rate is lost if the SI channel is changing rapidly, as it must then be estimated more frequently. Without a calibration period, a shorter channel coherence time results only in increased computational complexity due to the more frequent channel estimation, but none of the rate is lost.

Because of the more elaborate SI channel estimation procedure, the achievable instantaneous rate must now be written as a sum of two terms. The first term represents the half-duplex period during which the SI channels are estimated, while the second term represents the full-duplex period. Thus, the achievable instantaneous rate can be written as follows:
\begin{align}
	C_{c} &= \left(\frac{2N_c}{T_{coh} F_s} \right) \log_2 \left(1+\mathit{snr}\right)\nonumber\\
&+2\left(1-\frac{2N_c}{T_{coh} F_s}\right) \log_2\left(1+\mathit{sinr}_{c}\right) \text{,} \label{eq:cap_c}
\end{align}
where $N_c$ is the parameter estimation sample size used to estimate the SI channel during each coherence time period, $T_{coh}$ is the coherence time of the SI channel, $F_s$ is the sampling frequency, and $\mathit{sinr}_{c}$ is the achieved SINR at detector input in this scenario. Furthermore, for simplicity, it is assumed above that the thermal noise SNRs at each end's receiver, when in half-duplex mode, are identical (denoted by $\mathit{snr}$).

In Section~\ref{sec:simul}, numerical values for $C_{nc}$ and $C_{c}$ will be provided. Basically, they will quantify the trade-off between allocating resources for self-interference-free SI channel estimation and having a less accurate SI channel estimate due to the additional interference caused by the signal of interest. Based on the results, it is then possible to discuss whether it is more beneficial to perform the SI channel estimation while having a separate calibration period, or if it is better to do it during the actual reception.

\section{Waveform Simulations and Numerical Illustrations}
\label{sec:simul}

\begin{figure*}[!t]
\centering
\includegraphics[width=0.8\textwidth]{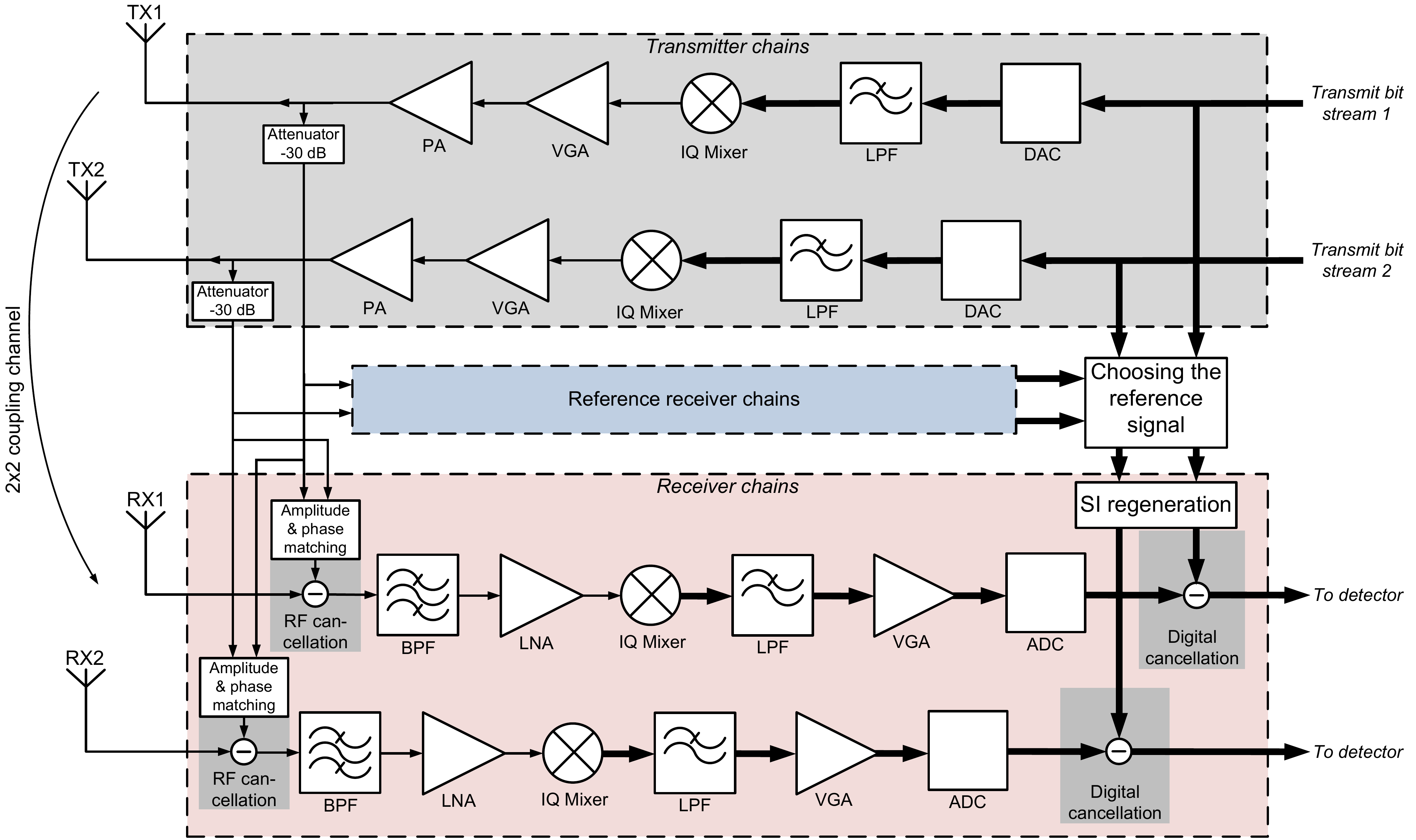}
\caption{A block diagram of the considered MIMO full-duplex transceiver, which has two alternative reference signal path options, for generality, for digital cancellation.}
\label{fig:block_diagram}
\end{figure*}

In the simulations, we model a 2x2 MIMO full-duplex transceiver, whose block diagram is presented in Fig.~\ref{fig:block_diagram}. Two possible reference signal paths for digital cancellation are shown in the block diagram, both of which are considered in the simulations. The reference receiver based digital cancellation is discussed in more detail in \cite{Korpi14}. In this paper, when simulating the case corresponding to linear digital cancellation, the reference signals are taken from the output of the transmitter chains, and in the widely-linear case the original transmitted samples are used as a reference. Note that, in the latter case, the circularity criterion of the reference signal is fulfilled, as required.

The parameters chosen for the transceiver are specified in Tables~\ref{table:system_parameters} and~\ref{table:parameters}, and the values have been chosen based on earlier literature and recent radio system specifications \cite{Parssinen99,Behzad07,Gu06,Jain11,Duarte10,LTE_specs}. Thus, even though the full-duplex transceiver was assumed to be linear in the above discussion, we make no such assumption in the actual simulations. Also note that in this analysis it is assumed that the receiver chain has been calibrated properly to achieve a fairly high image rejection ratio (IRR). This is a feasible assumption for a typical direct-conversion receiver \cite{Anttila08}.

In the simulations, least squares is used to determine the block wise SI channel estimate, as it is simple to implement and provides good performance. In addition, when assuming a linear model with Gaussian noise, it is also an efficient estimator \cite{Kay93}. For the linear signal model presented in Section~\ref{sec:linear_model}, the least squares channel estimate can be calculated as
\begin{align}
	\mathbf{\hat{h}}_i = (\mathbf{{X}}_{ref}^H \mathbf{{X}}_{ref} )^{-1}\mathbf{{X}}_{ref}^H \mathbf{y}_{i,ADC} \text{,} \label{eq:ch_est}
\end{align}
assuming $\mathbf{{X}}_{ref}$ has full column rank. For the widely-linear signal model presented in Section~\ref{sec:wl_model}, $\mathbf{{X}}_{ref}$ is replaced by $\mathbf{{X}}_{ref,WL}$ in the above expression.

 
\begin{table}[!t]
\renewcommand{\arraystretch}{1.3}
\caption{System level parameters of the proposed 2x2 MIMO full-duplex transceiver, alongside with the parameters specifying the utilized OFDM waveform.}
\label{table:system_parameters}
\centering
\begin{tabular}{|c||c|c|c||c|}
\cline{1-2} \cline{4-5}
\textbf{Parameter} & Value & & \textbf{Parameter (cont.)} & Value\\
\cline{1-2} \cline{4-5}
SNR target & 10 dB & & IRR (TX) & 25 dB\\
\cline{1-2} \cline{4-5}
Bandwidth & 12.5 MHz & & IRR (RX) & 60 dB\\
\cline{1-2} \cline{4-5}
RX noise figure & 4.1 dB & & Constellation & 16-QAM\\
\cline{1-2} \cline{4-5}
Sensitivity & -88.9 dBm & & Number of subcarriers & 64\\
\cline{1-2} \cline{4-5}  
RX signal power & -84.9 dBm & & Guard interval & 16 samples\\
\cline{1-2} \cline{4-5}
Transmit power & 10 dBm & & Sample length & 15.625 ns\\
\cline{1-2} \cline{4-5}
Antenna separation & 40 dB & & Symbol length & 4 $\mu$s\\
\cline{1-2} \cline{4-5}
RF cancellation & 30 dB & & Oversampling factor & 4\\
\cline{1-2} \cline{4-5} 
ADC bits & 12 & & Signal bandwidth & 12.5 MHz\\
\cline{1-2} \cline{4-5} 
PAPR & 10 dB & &  & \\
\cline{1-2} \cline{4-5}
\end{tabular}
\end{table}

\begin{table}[!t]
\renewcommand{\arraystretch}{1.3}
\caption{Parameters for the relevant components of the transmitter and receiver chains.}
\label{table:parameters}
\centering
\begin{tabular}{|c||c||c||c||c|}
\hline
\textbf{Component} & \textbf{Gain (dB)} & \textbf{IIP2 (dBm)} & \textbf{IIP3 (dBm)} & \textbf{NF (dB)}\\
\hline
PA (TX) & 27 & - & 15 & 5\\
\hline
LNA (RX) & 25 & - & 5 & 4.1\\
\hline
Mixer (RX)& 6 & 50 & 15 & 4\\
\hline
VGA (RX) & 0-69 & 50 & 20 & 4\\
\hline
\end{tabular}
\vspace{-2mm}
\end{table}


\subsection{Evaluating the Channel Estimator Variances}

To assess the validity of the above derivations regarding the estimator variance, we will first note that, based on \eqref{eq:N_req_lin}, the ratio between the parameter estimation sample sizes with and without calibration period should be $(\mathit{snr}+1)$, and it applies for both the linear and widely-linear signal models. The variances of the SI channel estimates are approximated based on the achieved SINRs in the simulations. Namely, it is assumed that if the SINRs are the same with and without the calibration period, then the variances of the channel estimates are also equal. This deduction is supported by \eqref{eq:y_adc_dc}, which illustrates that the power of the residual SI is directly dependent on the channel estimator variance. This type of an approach provides reliable results as long as the SINR is mainly limited by the accuracy of the SI channel estimate. However, when the accuracy of the estimator is sufficiently high and the SINR saturates, the estimator variance has only a small effect on the SINR. Thereby, the reliability of the above procedure decreases with very high values of $N$. Nevertheless, an important reason why we chose to do the comparison in terms of the SINRs is to show that the obtained equation actually provides accurate information about the overall performance of the MIMO full-duplex transceiver, and its applications are thereby not limited to assessing the estimator variances.

\begin{figure}[!t]
\centering
\includegraphics[width=\columnwidth]{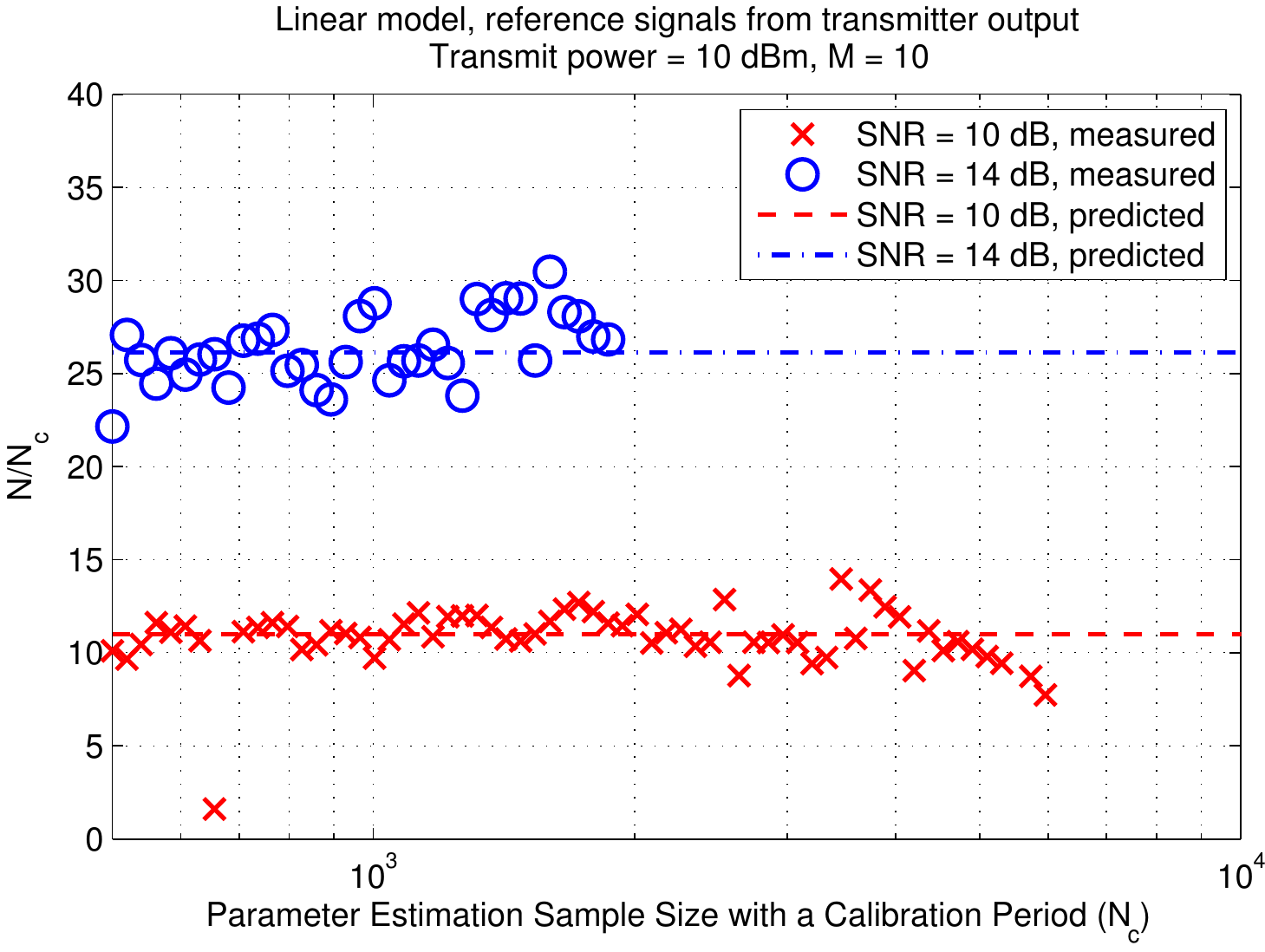}
\caption{Ratio between the parameter samples sizes with and without a calibration period, when linear digital cancellation is performed. The reference signals are taken from the output of the transmitter chains.}
\label{fig:Ns_lin}
\end{figure}

\begin{figure}[!t]
\centering
\includegraphics[width=\columnwidth]{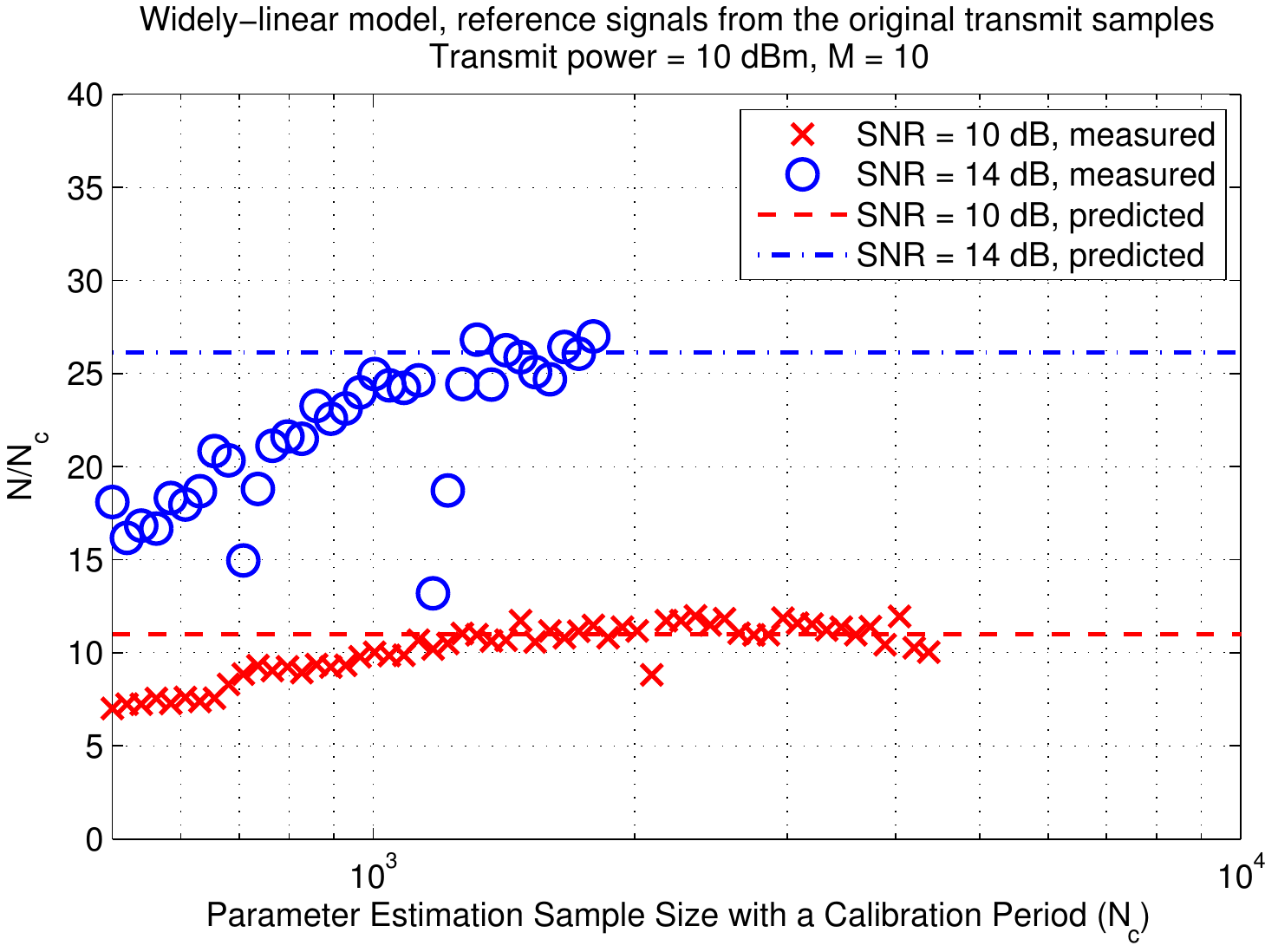}
\caption{Ratio between the parameter samples sizes with and without a calibration period, when widely-linear digital cancellation is performed. Now the original transmit samples are used as reference signals.}
\label{fig:Ns_wl}
\end{figure}

In Figs.~\ref{fig:Ns_lin} and~\ref{fig:Ns_wl} we have calculated the ratio $\frac{N}{N_c}$ with different values of $N_c$ for the linear and widely-linear signal models, respectively. In particular, we have given different values for $N_c$, determined the corresponding SINR achieved with a separate calibration period, and then measured how many samples are required to obtain a similar SINR without a calibration period. 

It can be observed from Fig.~\ref{fig:Ns_lin} that, with the linear model, where the reference signals are taken from the output of the transmitter chains, \eqref{eq:N_req_lin} provides remarkably accurate predictions with both of the considered SNR values, apart from few outliers. With a SNR of 14 dB at the detector input, the ratio could be measured for $N_c$ values of up to 2000 samples, and until that point the prediction is rather accurate. With the lower SNR, the value of $N_c$ could be increased up to 6000 samples, and again the prediction made by \eqref{eq:N_req_lin} regarding the ratio $\frac{N}{N_c}$ seems to be relatively accurate.

When considering the widely-linear model, where the original transmit samples are used as reference signals, Fig.~\ref{fig:Ns_wl} indicates that \eqref{eq:N_req_lin} provides accurate predictions only when $N_c \geq 1000$. Nevertheless, as long as this requirement holds, \eqref{eq:N_req_lin} seems to again provide accurate predictions regarding the parameter estimation sample size with and without a separate calibration period, regardless of the SNR. Thus, it can be concluded that the approximations made during the derivations in Sections~\ref{sec:linear_model} and~\ref{sec:wl_model} are valid, and the resulting relation between the parameter estimation sample sizes in \eqref{eq:N_req_lin} provides accurate predictions under wide circumstances, even when comparing the overall achieved SINRs.

\subsection{Achievable Rates under Practical Conditions}

Using the same transceiver model and parameters as above, let us now determine the achievable rates with and without a calibration period under practical conditions. For brevity, only widely-linear digital cancellation is considered in this context, as the essential results were observed to be largely similar also for linear digital cancellation.

\begin{figure}[!t]
\centering
\includegraphics[width=\columnwidth]{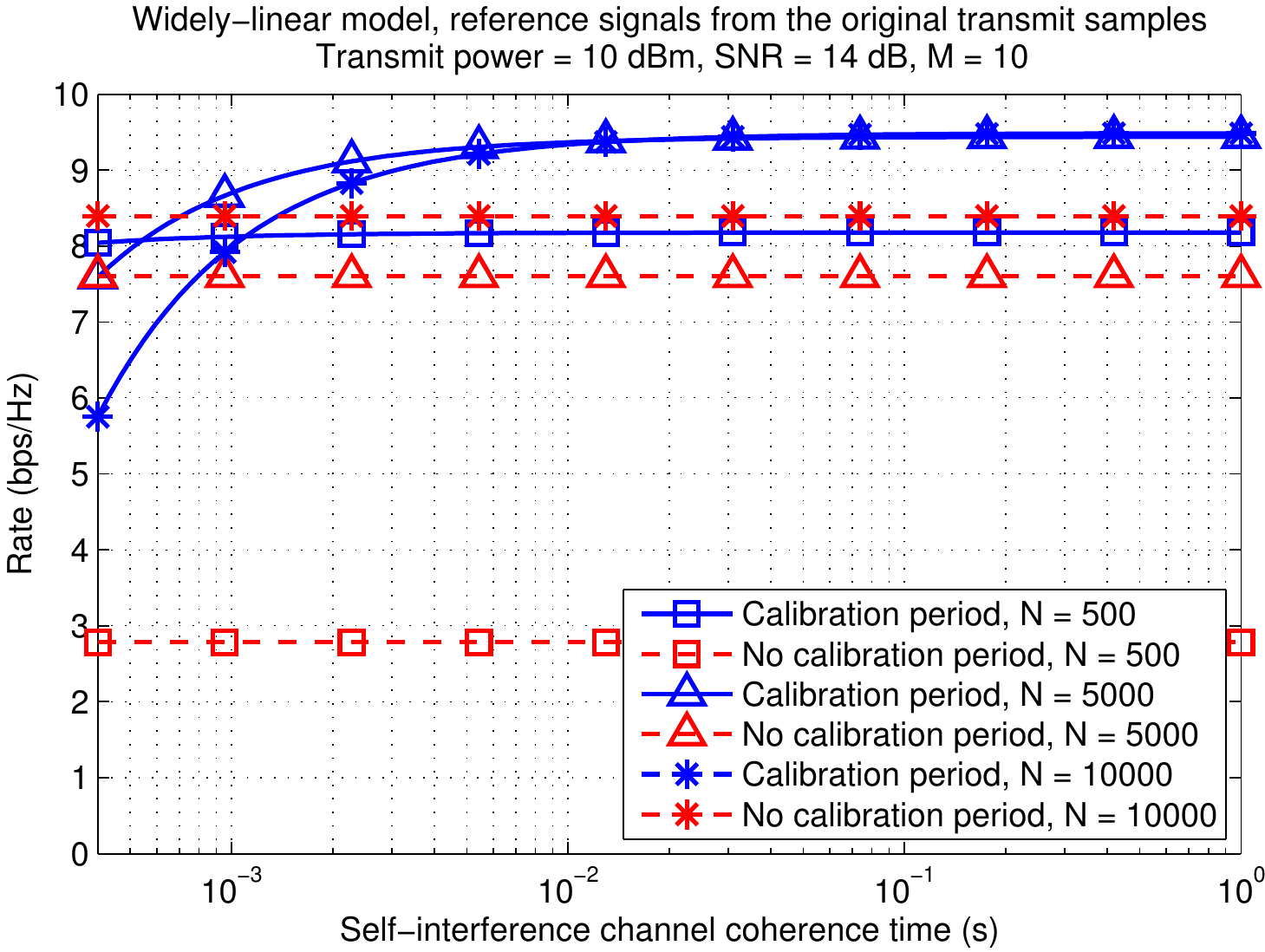}
\caption{The achievable average rates with and without a calibration period, with respect to the SI channel coherence time. The rates have been plotted for three different values of $N$ and for a given thermal noise SNR of 14 dB.}
\label{fig:coh_time}
\end{figure}

As illustrated by Fig.~\ref{fig:mac}, the coherence time of the SI channel is a crucial factor for the efficiency of a calibration period based full-duplex network. For this reason, Fig.~\ref{fig:coh_time} shows the achievable rates as a function of the channel coherence time, i.e., the interval between consecutive channel estimation cycles. The instantaneous rates have been calculated for three different parameter estimation sample sizes, using \eqref{eq:cap_nc} and \eqref{eq:cap_c} by determining the corresponding SINRs from the simulations, which are averaged over different SI channel realizations.

From Fig.~\ref{fig:coh_time} it can be observed that the achievable rate with a calibration period is directly proportional to the channel coherence time, as expected. Thus, with a very short channel coherence time, the rate is also relatively low. However, if the SI channel estimation is done during full-duplex operation, the achievable rate is not affected by the channel coherence time. In fact, only the computational complexity of the digital cancellation is increased because the SI channel must be estimated more frequently. In Fig.~\ref{fig:coh_time} this is shown by the fact that the rates without a calibration period are constant with respect to the coherence time.\footnote{Notice that all the results are here for given received signal of interest thermal noise SNR of 14 dB, while in the final paper, we will also provide corresponding average rate results where the fading of the signal of interest is explicitly included in the averaging.}

When comparing the two scenarios, it can be concluded that performing SI channel estimation without a calibration period is the preferable option if the SI channel changes very rapidly. With a sufficiently large parameter estimation sample size, using a calibration period becomes the better option only if the channel coherence time is in the order of $10^{-3}$ seconds. After this, the burden of the calibration period is less significant than the benefit of having a more accurate SI channel estimate 

When investigating the effect of $N$ on the achievable rates, it can be observed that, with the very small parameter estimation sample size of 500, having a calibration period provides approximately a threefold increase in the achievable rate, regardless of the channel coherence time. This is explained by the high variance of the SI channel estimate when there is no separate calibration period, since 500 samples is not enough to produce a sufficiently accurate estimate in that case.


However, when assuming more practical values for $N$, it becomes evident that having a separate calibration period does not improve the rate that drastically. Overall, it performs worse with the shorter channel coherence times, while providing a moderate performance gain under more static channel environments. If $N \geq 5000$, having a calibration period provides a 1--2 bps/Hz increase in the rate with the longer channel coherence times, depending on the parameter estimation sample size. Thereby, when considering that performing the SI channel estimation during a calibration period requires quite a bit of complexity in the medium access control (MAC) layer, it seems that having no calibration period is indeed a viable option from the achievable rate perspective.

\subsection{Further Discussion}

In this paper, it is assumed that the SI channel is estimated once during each channel coherence time period. However, without separate a calibration period, it would be possible to estimate the SI channel more frequently, and/or with a longer data block, thereby having a more accurate estimate at each time instant since, in practice, the SI channel would vary slightly also during the coherence time. This would of course require an increased amount of computational resources, but would also provide a performance gain in comparison to estimating the SI channel only once in each coherence time period.

Another possible option to utilize the freedom of having no fixed calibration period would be to use adaptive algorithms to constantly track the SI channel. Using, for example, recursive least squares (RLS) would provide an opportunity to continuosly estimate the SI channel while consuming relatively little computational resources. The results derived here can be used, e.g., to approximate the variance of the RLS based channel estimator with finite asymptotic sample length of the form $1/(1-\lambda)$ where $\lambda$ denotes the RLS forgetting factor. Thus, even though the rate achieved without a calibration period is typically slightly lower than that achieved with a calibration period, there are also other factors that affect the final achievable performance.

\section{Conclusion}
\label{sec:conc}

In this paper we have analyzed self-interference channel estimation with and without a separate calibration period, in terms of estimator variance lower bound. We derived an equation quantifying the increase in the required parameter estimation sample size when estimating the self-interference channel during normal operation, i.e., when there is also a received signal of interest included in the total signal, which acts as noise from the estimation perspective. It was shown that the derived equation holds for both linear and widely-linear digital cancellation. Its validity for a practical 2x2 MIMO full-duplex transceiver was also confirmed with full waveform simulations. In addition, we analyzed the achievable rates with and without a periodical calibration period to determine whether it is preferable to have the additional overhead introduced by it or a less accurate self-interference channel estimate. It was shown that the overall achievable rate without a separate calibration period was either higher or slightly lower than that achieved with a separate calibration period, depending on the self-interference channel coherence time. Thus, under rapidly changing self-interference channel conditions, performing the self-interference channel estimation without a calibration period is a viable option. 


\bibliographystyle{./IEEEtran}
\bibliography{../IEEEabrv,../IEEEref}

\end{document}